\documentclass{article}

\usepackage[final, main]{neurips_2025}

\workshoptitle{AI for Science workshop (NeurIPS 2025)}

\usepackage[utf8]{inputenc} 
\usepackage[T1]{fontenc}    
\usepackage{hyperref}       
\usepackage{url}            
\usepackage{booktabs}       
\usepackage{amsfonts}       
\usepackage{nicefrac}       
\usepackage{microtype}      
\usepackage{xcolor}         
\usepackage{graphicx}
\usepackage{float}

\title{A Synthesizability-Guided Pipeline for Materials Discovery}

\usepackage{authblk} 

\definecolor{brandred}{HTML}{C62828}
\newcommand{\red}[1]{\textcolor{black}{#1}}

\author[2]{Thorben Prein}
\author[1]{Willis O'Leary}
\author[1]{Aikaterini Flessa Savvidou}
\author[1]{Elchaïma Bourneix}
\author[1]{Joonatan E. M. Laulainen}

\affil[1]{Altrove, 25 Rue de Ponthieu, Paris, France}
\affil[2]{Technische Universität München, Garching b. München, Germany}

\begin{document}
\maketitle

\begin{abstract}
Computational materials discovery relies on the generation of plausible crystal structures. The plausibility is typically judged through density functional theory methods which, while typically accurate at zero Kelvin, often favor low-energy structures that are not experimentally accessible. We develop a combined compositional and structural synthesizability score which provides an accurate way of predicting which compounds can actually be synthesized in a laboratory. We use it to evaluate non-synthesized structures from the Materials Project, GNoME, and Alexandria, and identified several hundred highly synthesizable candidates. We then predict synthesis pathways, conduct corresponding experiments, and characterize the products across 16 targets, \textbf{successfully synthesizing 7 of 16}. The entire experimental process was completed in only three days. Our results highlight omissions in lists of known synthesized structures, deliver insights into the practical utility of current materials databases, and showcase the central role synthesizability prediction can play in materials discovery.
  
\end{abstract}

\section{Introduction}
\label{sec:intro}

Discovery of new inorganic materials is a central goal of solid-state chemistry and, when achieved, can usher in enormous scientific and technological advancements. The development of machine-learning–accelerated, ultra-fast \textit{in-silico} screening \citep{batzner20223, chen2022universal, wood2025family, rhodes2025orb} has unlocked vast databases of predicted candidate structures. \red{Efforts} such as the Materials Project \citep{horton2025accelerated}, GNoME \citep{merchant2023scaling}, and Alexandria \citep{schmidt2023machine} contain structures predicted using active learning and a mix of composition- and structure-based models to propose candidate structures, which are subsequently assessed by density functional theory (DFT) simulations. Analogous to virtual libraries in drug discovery \citep{zdrazil2024chembl, tingle2023zinc}, these resources provide initial property estimates alongside full crystallographic descriptions of putative phases.

The number of proposed inorganic crystals now exceeds the number of experimentally synthesized compounds by more than an order of magnitude \citep{belsky2002new}. The current challenge is to determine which of these predicted materials can be fabricated. This has historically been addressed by computing convex-hull stabilities with DFT \citep{lee2022machine}. While this approach constitutes a useful first filter, it typically overlooks finite-temperature effects, namely entropic and kinetic factors, that govern synthetic accessibility \citep{bartel2022review}. As a result, it has become increasingly difficult to distinguish purported stable structures from truly synthesizable ones. For example, the Materials Project lists 21 SiO$_2$ structures within 0.01 eV of the convex hull (as of August 2025). The second most common phase SiO$_2$, cristobalite ($\beta$-quartz), is not among these 21. Overall, there is a pressing need for accurate synthesizability assessments to efficiently steer scientists and engineers toward compounds that are readily accessible in the laboratory \citep{park2025closing}.

Given the abundance of predicted near-stable crystal structures, procedures are needed to decide which candidates to attempt to synthesize and under which conditions these syntheses should be carried out (process parameters). So far, several data-driven approaches have been introduced, including graph-based structural models and composition-based models \citep{lee2022machine,antoniuk2023predicting, song2025accurate, davariashtiyani2021predicting, zhu2023predicting, gleaves2023materials, amariamir2025syncotrain}. These approaches consider composition and structure in isolation and, to date, have not convincingly demonstrated that they can drive the experimental synthesis of novel compounds. Here, we propose and implement a unified, synthesis-aware prioritization framework that integrates complementary signals from composition and crystal structure. Composition signals may be governed by elemental chemistry, precursor availability, redox and volatility constraints, whereas structural signals capture local coordination, motif stability, and packing. Our unified model demonstrates state-of-the-art performance. We apply our framework to more than 4.4 million simulated crystal structures, yielding 24 candidates predicted to be highly synthesizable. Subsequently, we employ neural models trained on literature-mined synthesis recipes to predict process parameters and subsequently carry out syntheses on the proposed candidates in an automated solid-state laboratory. Of the 16 samples we successfully characterized, seven matched the target structure,  including one completely novel and one previously unreported structure.

\section{Methods}

\begin{figure}[H]
    \centering
    \includegraphics[width=0.99\linewidth]{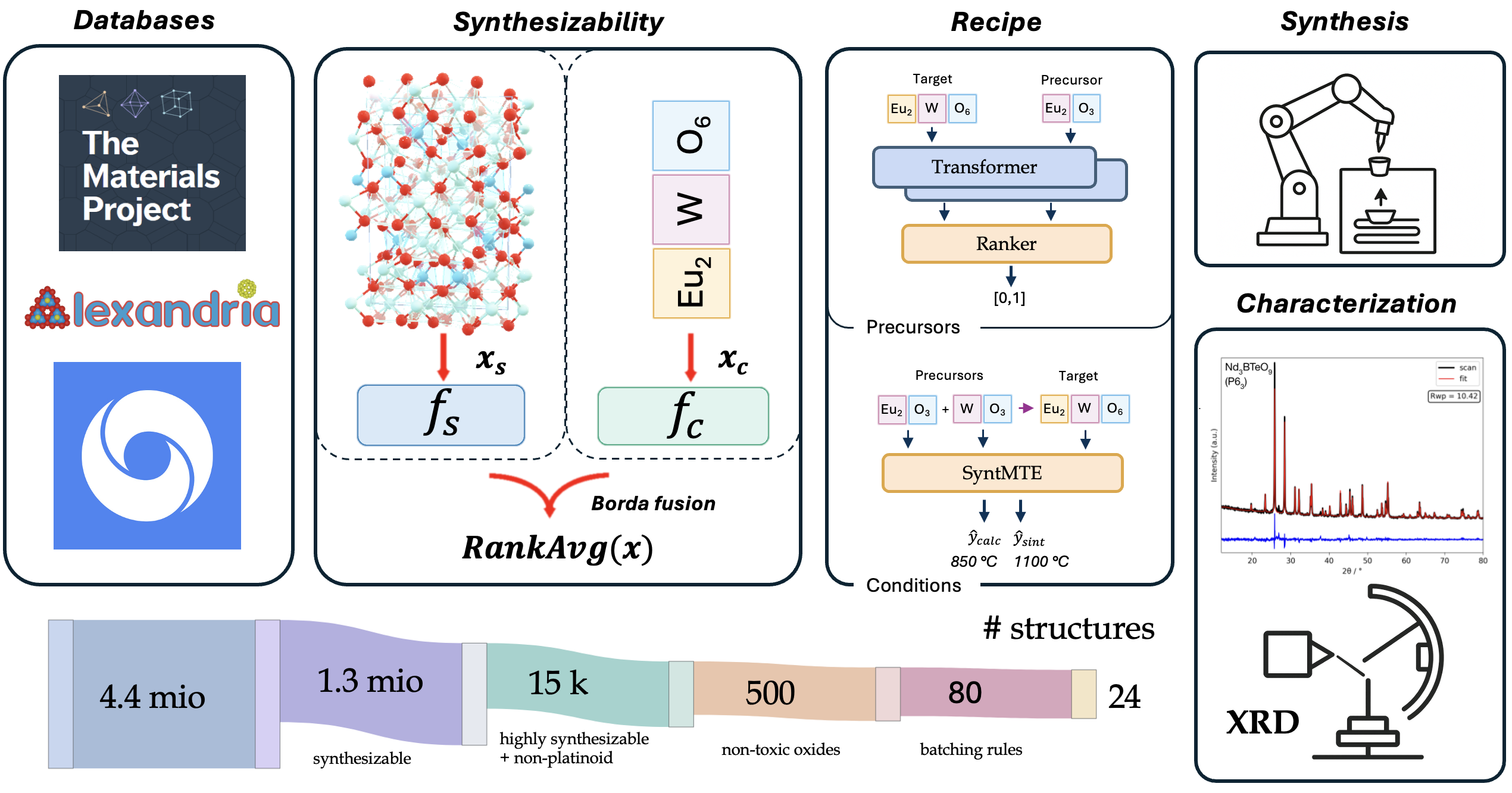} 
    \caption{Our pipeline for selecting synthesizable candidates, predicting their recipe, carrying out the synthesis, and characterizing their structure. 
    }
    \label{fig:rank_fusion}
\end{figure}

Fig. \ref{fig:rank_fusion} outlines our synthesizability-guided pipeline. We first screened a pool of 4.4 million computational structures, of which 1.3 million were calculated to be synthesizable (detailed distributions Fig, \ref{fig:pca} and Fig. \ref{fig:synthesizability_by_compound_class}). We then focused only on highly synthesizable (0.95 rank-average per our synthesizability score) materials and removed compounds containing platinoid group elements, yielding approximately 15,000 candidates. Finally, we reduced the candidates by removing non-oxides and toxic compounds, yielding approximately 500 final structures. We applied retrosynthetic planning on these prioritized structures to generate feasible routes, of which a subset (see Section \ref{subsec:experiment}) was then executed in a high-throughput laboratory platform. The resulting products were verified automatically by X-ray diffraction (XRD) (see Appendix for details).

\subsection{Synthesizability Model}

We define synthesizability as the probability that a compound can be prepared in the laboratory using currently available synthetic methods. Machine-learning efforts to predict this outcome largely fall into two families: (i) composition-only models that operate on stoichiometry or engineered composition descriptors \citep{lee2022machine,antoniuk2023predicting}; and (ii) structure-aware models that leverage crystal-structure graphs \citep{song2025accurate,davariashtiyani2021predicting,zhu2023predicting,gleaves2023materials,amariamir2025syncotrain}.

\paragraph{Problem Formulation}
Let each candidate material be represented by its composition $x_c$ and a relaxed crystal structure $x_s$. We denote the input as $x=(x_c,x_s)$ and the label $y\in\{0,1\}$ indicating whether the compound has been experimentally synthesized. The goal is to learn a score $s(x)\in[0,1]$ that estimates a compound's synthesizability.

\paragraph{Data Curation}

We built our training dataset from the Materials Project \citep{jain2013commentary}. This ensures consistency between composition and structure in order to obtain supervision without inheriting artifacts common in experimental entries encountered e.g. in the Inorganic Crystal Structure Database (ICSD) \citep{hellenbrandt2004inorganic}, such as non-stoichiometry, dopants or partial occupancies. We assigned labels according to the Materials Project's ``theoretical'' field, a flag which corresponds to whether ICSD entries exist for a given structure. We labeled a composition as unsynthesizable ($y=0$) if all polymorphs of that composition are flagged as theoretical. Conversely, we labeled a composition as synthesizable ($y=1$) if there exists any polymorph that is not flagged as theoretical.  The final dataset comprised 49,318 synthesizable compositions and 129,306 unsynthesizable compositions, stratified into train/validation/test splits.

\paragraph{Model}
Our model integrates complementary signals from composition and structure via two encoders, \red{learning a binary classification task}:
\[
\mathbf{z}_c = f_c(x_c;\,\theta_c),\qquad
\mathbf{z}_s = f_s(x_s;\,\theta_s),
\]
where $f_c$ is a fine-tuned compositional MTEncoder transformer \citep{prein2023mtencoder} and $f_s$ is a graph neural network fine-tuned from the JMP model \citep{shoghi2023molecules} on the crystal structure $x_s$. Both encoders are extensively pretrained and each feeds a small MLP head that outputs a separate synthesizability score. During training, we fine-tune all parameters end-to-end on an NVIDIA H200 cluster. We minimize binary cross-entropy with early stopping on validation AUPRC.

\paragraph{Usage in Screening}
At inference, each model outputs a synthesizability probability for candidate \(x_i\).
To obtain an enhanced ranking across candidates, we aggregate both predictions via a
rank-average ensemble (Borda fusion). Specifically, we convert probabilities to ranks and define
\[
\mathrm{RankAvg}(i)
=\frac{1}{2N}\sum_{m\in\{c,s\}}\left(1+\sum_{j=1}^{N}\mathbf{1}\!\big[s_m(j)<s_m(i)\big]\right).
\]
Here \(i\in\{1,\dots,N\}\) indexes candidates. \(N\) is the total number of candidates in the screening pool. The index \(m\in\{c,s\}\) refers to the composition and structure models, and \(s_m(i)\in[0,1]\) is the synthesizability probability predicted by model \(m\) for candidate \(i\). By construction, \(\mathrm{RankAvg}(i)\in[1/N,1]\), with larger values indicating greater predicted synthesizability. Consequently, we rank candidates by \(\mathrm{RankAvg}\) rather than applying probability thresholds.

\paragraph{Synthesis Planning}
After identifying high-priority candidates with the rank-based screening, we generated synthesis recipes in two stages. First, we applied Retro-Rank-In \citep{prein2025retro}, a precursor-suggestion model, to produce a ranked list of viable solid-state precursors for each target. Second, we selected the top-ranked precursor pairs and use SyntMTE \citep{prein2025language} to predict the calcination temperature required to form the target phase. We then balanced the reaction and compute the corresponding precursor quantities. Both models are trained on literature-mined corpora of solid-state synthesis \citep{kononova2019text}.

\begin{figure}
    \centering
    \includegraphics[width=\linewidth]{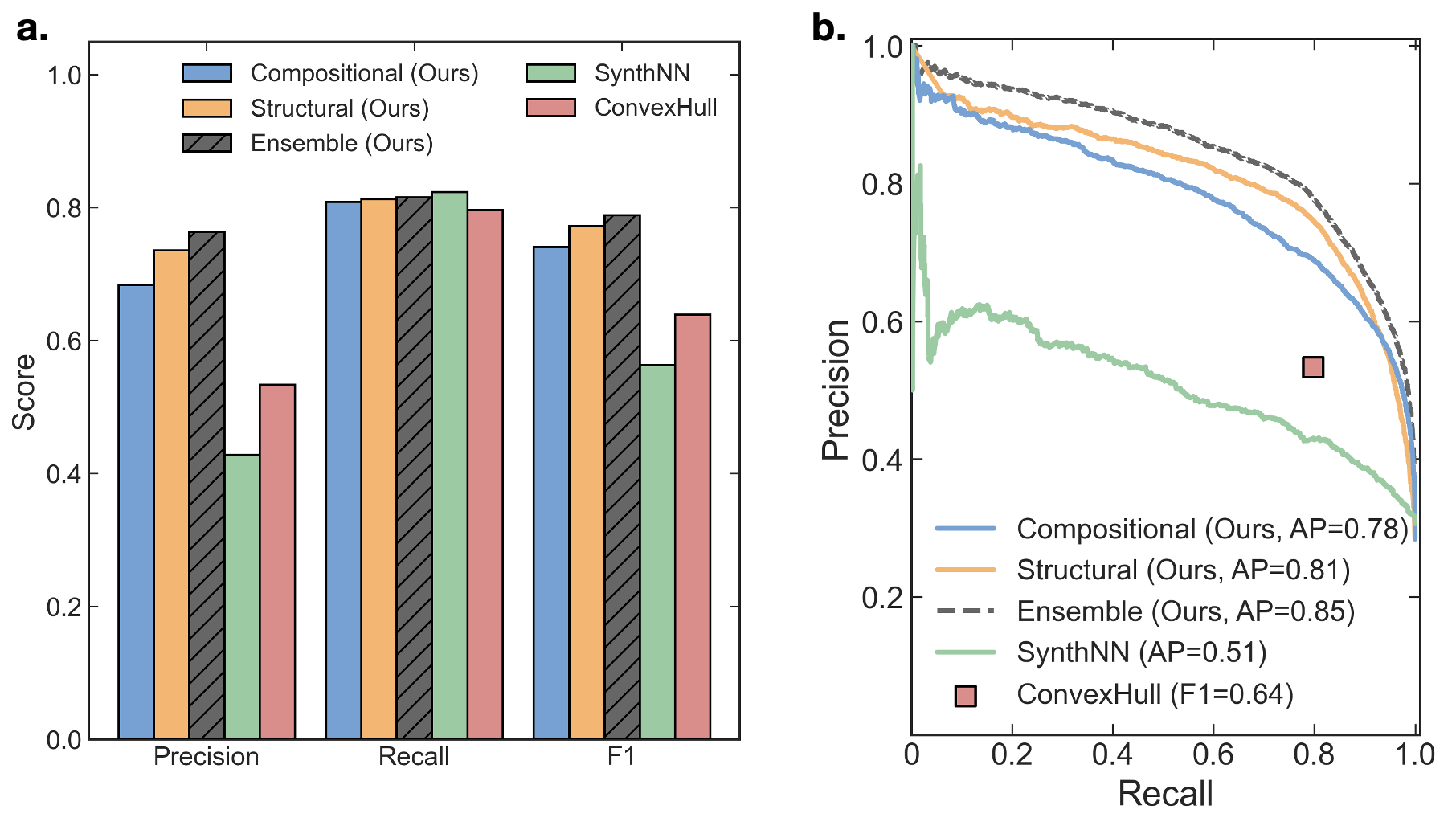} 
    \caption{Test-set performance of the compositional, structural, and ensemble models. \textbf{(a)} Precision, recall, and F1 scores for identifying the target "theoretical" labels in the materials project. \textbf{(b)} ROC curves (AUC shown in the legend). The convex-hull baseline classifies compounds within \(50\,\mathrm{meV}\) above the hull as synthesizable.}
    \label{fig:performance}
\end{figure}

\subsection{Experimental Synthesis and Characterization}
\label{subsec:experiment}

Of the ca. 500 candidates \red{shown in Fig. \ref{fig:rank_fusion}}, we selected the experimental candidates by a combination of a web-searching LLM to judge whether each candidate was likely to have been synthesized before, followed by an expert judgment to remove targets with unrealistic oxidation states given an oxygenated furnace (e.g. BaTe$_2$O) and those with common formulas (e.g. DyO$_2$)  as these compounds are likely to be well-explored. This yielded a final list of 80 targets.

Next, based on recipe similarity, we automatically selected 24 targets across two batches of 12, such that all 12 experiments in one batch could be carried out in the furnace at the same time. Finally, the samples were weighed, ground and calcined in a Thermo Scientific Thermolyne Benchtop Muffle Furnace in our high-throughput laboratory (see Appendix for further experimental details). Eight of the 24 samples strongly bonded to their crucibles during synthesis and therefore could not be subjected to further characterization. The remaining 16 samples could be milled after synthesis and were subsequently characterized using a Malvern Panalytical Aeris benchtop X-ray diffractometer. The X-ray spectra were automatically analyzed, then expert analysis was used to confirm the presence of the target compared to existing phases.

\section{Results}

\subsection{Model Performance}

We benchmark our approach against two baselines: SynthNN, a recent structure-aware synthesizability model \cite{antoniuk2023predicting}, and a convex-hull classifier that labels compounds as synthesizable when their Materials Project energy above the hull is \(\leq 50\,\mathrm{meV}/\mathrm{atom}\). While both baselines attain competitive recall, they suffer from markedly lower precision than our method, which causes them to over-prioritize many compositions that are unlikely to be synthesizable (Fig.~\ref{fig:performance}). Specifically, SynthNN achieves \(R=0.823\) but only \(P=0.428\), and the convex-hull heuristic reaches \(R=0.797\) with \(P=0.534\).

To isolate the contributions inside our ensemble approach, we compare the compositional and the structural part with the final rank–average ensemble. As summarized in Fig.~\ref{fig:performance}\,(a,b), the structural model improves upon the compositional baseline across all metrics \(P=0.736\) vs.\ \(0.684\), \(R=0.813\) vs.\ \(0.809\), and \(F_1=0.773\) vs.\ \(0.741\) and the ensemble performs best overall, reaching \(P=0.764\), \(R=0.816\), and \(F_1=0.789\). Recall remains tightly clustered across our three variants, indicating that the bulk of improvement comes from higher precision. Since our final ensemble ranking lacks a meaningful global threshold, we calibrate each model’s threshold to maximize its F1 score to ensure comparability.  

We hypothesize that the compositional model’s competitiveness reflects that much of the synthesizability signal resides in elemental chemistry rather than exact atomic geometry, charge balance, valence matching, redox windows, and volatility constraints are inherently compositional. These signals recover many true positives from the unlabeled pool, boosting recall, but also inflate false positives for stoichiometries that appear feasible yet cannot form stable coordination networks. By encoding local coordination and connectivity, the structural model suppresses precisely these spurious cases, recovering precision with minimal cost in recall. The final rank–average ensemble inherits the compositional coverage while recapturing most of the lost precision, yielding the best overall balance.

\paragraph{Case Illustrations}
Let \(s_{\mathrm{comp}}\) denote the compositional prediction and \(s_{\mathrm{struct}}\) the structural prediction for the \emph{same} candidate structure.

\begin{itemize}

  \item \emph{N\textsubscript{2}}:
  \(s_{\mathrm{comp}}=0.709\), \(s_{\mathrm{struct}}=1.10\times10^{-4}\).
  While the composition looks plausible in isolation, with nitrogen-rich formulas occurring near many synthesized nitrides, the candidate is a molecular solid likely requiring non-ambient conditions. The structure encoder down-weights such graphs because they lack robust inorganic connectivity under our solid-state synthesis domain.

  \item \emph{Cu\textsubscript{6}Te\textsubscript{2}Mo\textsubscript{2}H\textsubscript{2}Cl\textsubscript{4}O\textsubscript{15}}:
  \(s_{\mathrm{comp}}=0.858\), \(s_{\mathrm{struct}}=1.66\times10^{-3}\).
  The composition sits near known oxyhalides and oxychalcogenides, leading to high \(s_{\mathrm{comp}}\), but the proposed structure mixes competing anions and volatile species that are difficult to stabilize in a single phase via high-temperature calcination. The structure model captures this through unfavorable coordination and packing cues, yielding a low \(s_{\mathrm{struct}}\).
\end{itemize}

Because our supervision stems from positive-unlabeled (PU) entries with reports of synthesized materials being treated as positives, while all others are unlabeled rather than true negatives, the precision values above should be viewed as conservative.
In PU settings, many predicted positives inside the unlabeled pool are plausible true positives, thus the observed increases in precision from structure and ensembling represent lower bounds on their practical utility.
Conversely, recall is computed on labeled positives and is therefore less sensitive to unlabeled contamination, explaining the narrow spread in recall across models.

\subsection{Experimental Results}

\begin{figure}
    \centering
    \includegraphics[width=\linewidth]{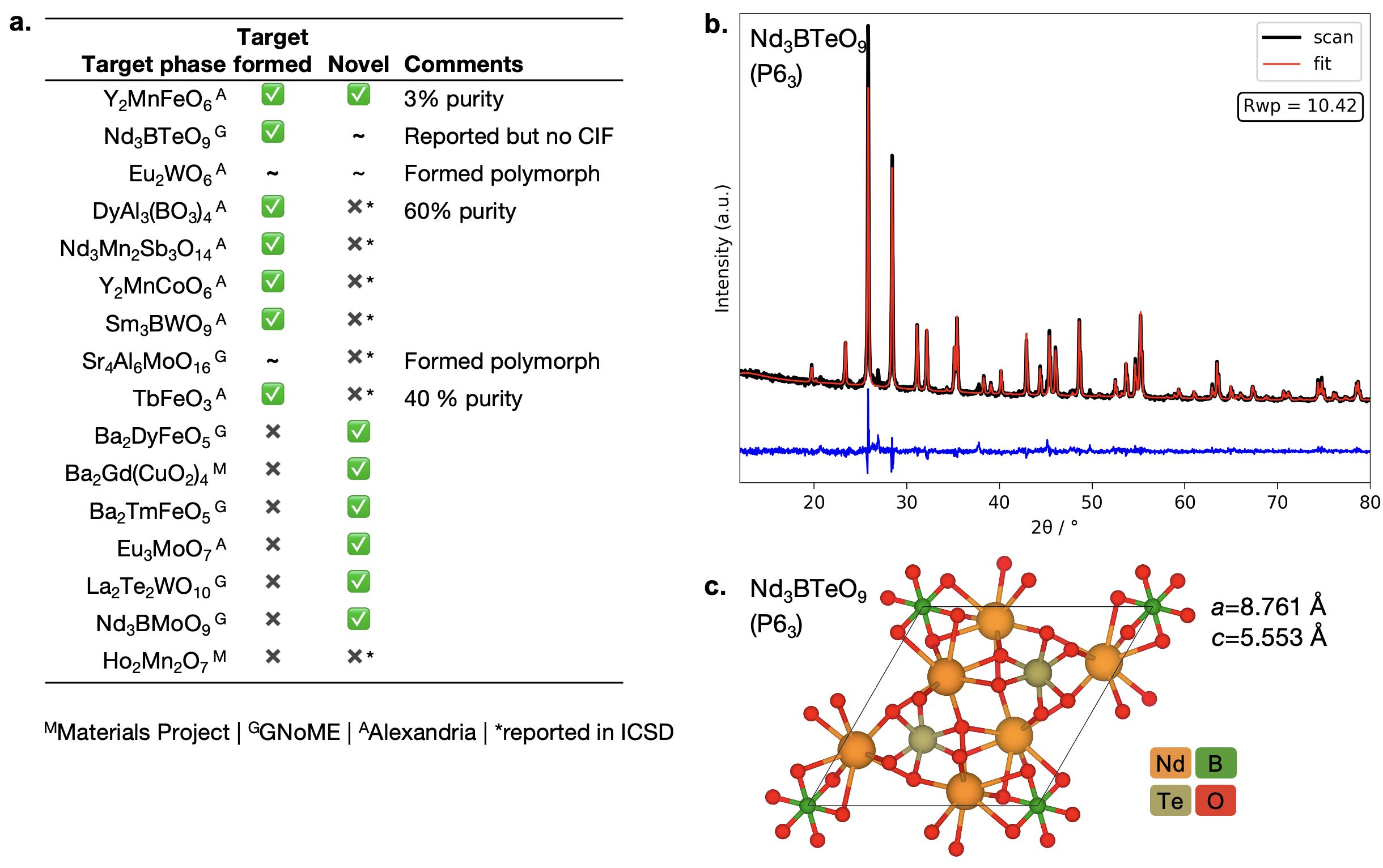}
    \caption{\textbf{a.} Identified highly synthesizable ($\mathrm{RankAvg}>0.95$) targets and the results of performed experiments.  For Eu$_2$WO$_6$, a non-target polymorph was synthesized that had the same space group as a known structure but with significantly different lattice parameters. For Sr$_4$Al$_6$MoO$_{16}$, a compound with the target composition was synthesized. However, due to low purity we could not rule out formation of one of the three known polymorphs. Several compounds, while listed as theoretical in Materials Project, do have similar structures reported, mostly in the ICSD.  \textbf{b.} XRD scan, fit, and difference for Nd$_3$BTeO$_9$. A second unknown phase is present and accounts for <5\% of total intensity.  \textbf{c.} Refined hexagonal unit cell for Nd$_3$BTeO$_9$. The unit cell is hexagonal with the lattice parameters $a=8.761$ Å and $c=5.553$ Å}
    \label{fig:xrd}
\end{figure}

 The targets, synthesis results, along with an example XRD spectrum and structure, are provided in Fig. ~\ref{fig:xrd}.  Of the 16 characterized samples, we successfully synthesized the target seven times, (44\%), synthesized a predicted or known polymorphs two times (12\%), and failed to synthesize the target 7 times (44\%) (Fig. ~\ref{fig:xrd}a). Of the nine syntheses where either the target or a polymorph was synthesized, three were subsequently identified to be nearly identical to a structure already in the Materials Project (and ICSD) that is listed as synthesized. These structures were therefore effectively present in the training data. Of the remaining six, a similar structure was present in the ICSD for three compounds, although these compounds were erroneously classed as theoretical in the Materials Project database. 
 
 We synthesized three compounds with no known structure reported in ICSD: 
 \begin{itemize}
    \item Y$_2$MnFeO$_6$ (Alexandria): Three polymorphs with the composition have been synthesized, while one theoretical structure is available both in Materials Project and Alexandria. We detected our target polymorph at a low (3.3\%) concentration in addition to observing a phase mixture of the other two known polymorphs (see Appendix for XRD fit).
     \item Nd$_3$BTeO$_9$ (GNoME): We found a paper listing its synthesis but without a CIF file \citep{zhou2023large}. Our refined structure is isostructural to Bi$_3$BTeO$_9$, with a P6$_3$ space group and $a=8.761$ Å and $c=5.553$ Å. We achieved near full purity, with $<5$\% of intensity accounted for by an unknown phase that we could not identify.
     \item Eu$_2$WO$_6$ (Alexandria): A CIF file for this structure exists in the ICSD, but with significantly (ca. 5\%) different lattice parameters. The observed orthorhombic polymorph was different than the original target and was formed with ca. 88\% purity.
 \end{itemize}

\section{Discussion}

\subsection{Synthesis of Novel Compounds}

The successful one-shot synthesis of (a) novel compounds and (b) known compounds excluded from our training set highlights the promise of our approach as well as the broader value of machine-learning-based synthesizability screening methods. Given current trends, we expect the number and diversity of predicted inorganic structures to continue to grow. We expect many of these structures to be stable according to DFT-derived metrics but ultimately unsynthesizable due to high kinetic barriers or stability only under unusual or wholly impractical conditions. With our approach, we enjoyed a 44\% success rate in selecting synthesizable structures from millions of possible candidates. Presumably, some candidates in the remaining 56\% may be synthesizable under different conditions or through the use of alternative synthesis methods. 

Still, our successes carry with them strong caveats. First, most of the successful synthesis candidates were isostructural to other known compounds. This might reflect our greedy selection of candidates with only the highest synthesizability probabilities. It may also arise from data limitations and the challenges to build models that generalize to exotic, never-before-seen structures. Furthermore, many of our successful syntheses were in fact not novel compounds, despite being marked as purely theoretical in Materials Project. Clearly, our automated screening to find unsynthesized compounds can be significantly improved. Although our results are encouraging as far as model performance is concerned, it also motivates efforts to provide more comprehensive and accurate data for training synthesizability models.

Finally, we would like to emphasize the importance of both the structural and compositional components of the model. In preliminary work, we attempted the synthesis of nine candidates selected using a composition-only model but formed none of the targets. Upon evaluating the candidates, we noted that the majority were unrealistic polymorphs of well-established compounds. In addition, we observed a  tendency to up-rank materials with complex ordering of transition metals (e.g. ScTa$_5$Nb$_6$Tc$_{12}$). These structures are rarely experimentally achieved due to entropic effects but tend to be overrepresented in DFT-derived materials datasets. These findings motivated trials with a purely structure-based model. The use of a structure-based model eliminated the aforementioned candidates but favored compounds with unrealistic chemistries (e.g.  AsH$_4$F$_7$). In the end, synthesizing the complementary signals of composition-based and structure-based  models led to superior overall performance in predicting synthesizability, \red{in this context, all of the candidates we tried in the lab achieved high scores in both modalities.} 

\begin{figure}[h!]
    \centering
    \includegraphics[width=\linewidth]{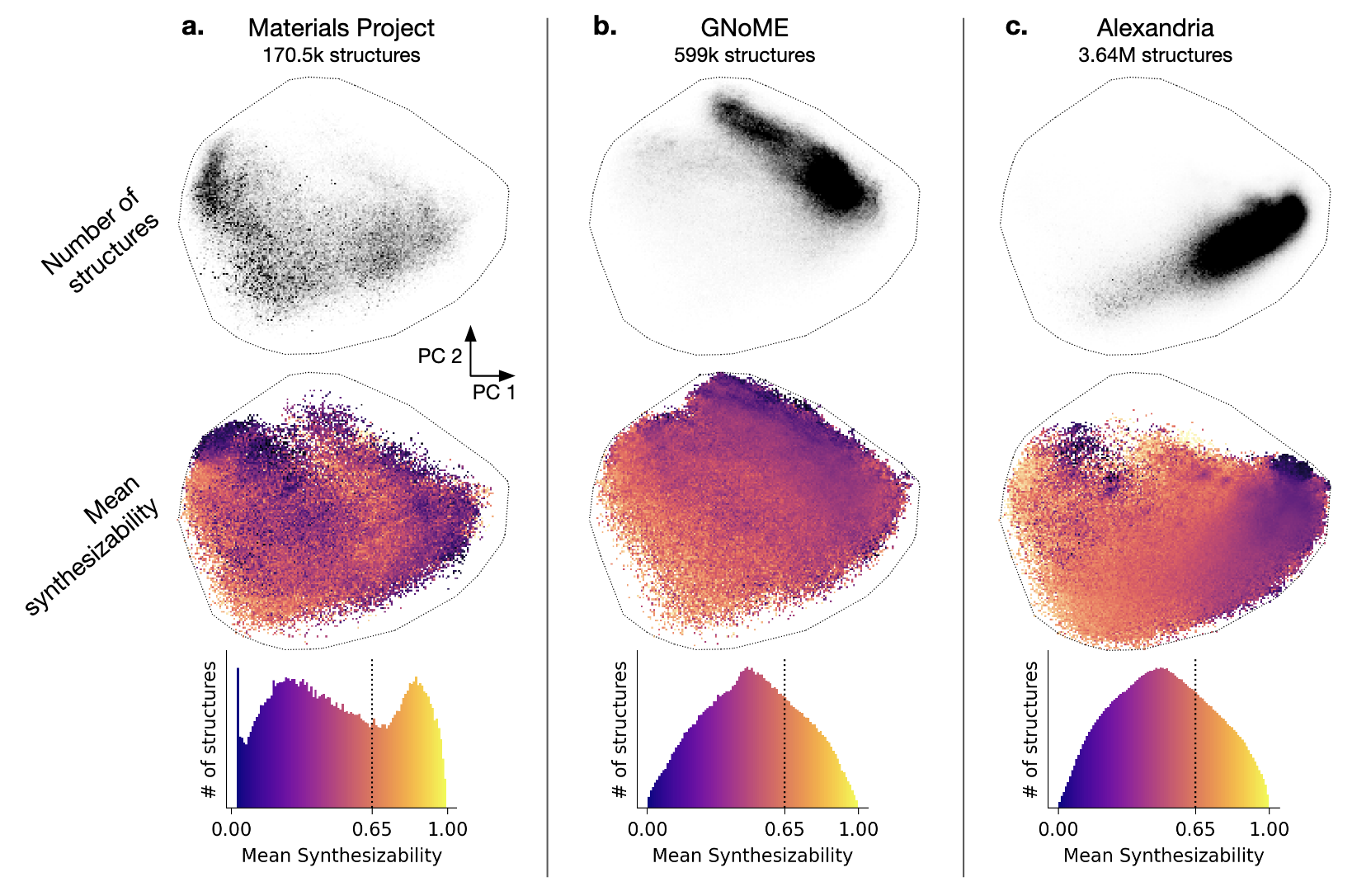}
    \caption{Joint principal component analyses on MTEncoder embeddings \red{shown in the same latent space} for  \textbf{a.} Materials Project,  \textbf{b.} GNoME, and  \textbf{c.} Alexandria. Top row: density of structures by embedding. Middle row: average synthesizability by embedding. Bottom row: distribution of rank-average synthesizabilities within each dataset. }
    \label{fig:pca}
\end{figure}

\subsection{Dataset-level Insights}

With a robust score for synthesizability in hand, we can better understand the distribution of synthesizabilities for the roughly 4.4 million materials available in Materials Project, GNoME, and Alexandria. To do this, we generated MTEncoder \citep{prein2023mtencoder} embeddings for all compositions contained within these datasets then carried out principal component analysis (PCA). Dataset metrics are plotted with respect to the first two principal components in Fig. \ref{fig:pca}. We find that these datasets cover quite distinct regions in principal component space. Materials Project contains a relatively diverse set of materials, distributed across the space, while the materials contained in GNoME and Alexandria are concentrated in comparatively narrow regions. 

For each database, we computed mean synthesizability across principal component space. Certain regions in Materials Project boast fairly high average synthesizability while others do not. The existence of ``patches'' of synthesizability is unsurprising, as we would expect a complex relationship between principal components and any material property. For GNoME and Alexandria, we observe a roughly inverse relationship between the number of materials and mean synthesizability in a given region. A number of possible factors could give rise to this trend. One intuitive reason is that, in the drive to explore more diverse structures and compositions, the GNoME and Alexandria projects have generated less synthesizable materials. For example, compounds containing a greater number of elements are less likely to have been synthesized, on average. In addition, the simplest structures and compositions have likely already been synthesized, and therefore are not present in GNoME or Alexandria. Finally, this trend may simply reflect differences in the data distribution between Materials Project -- the source of our positive labels -- and other databases. Overall, this highlights the need to critically judge predicted crystal structures for synthesizability, especially as the total number of predicted structures continues to grow.

\subsection{Data Limitations}

As noted above, data limitations, specifically a lack of a fully unified and accurate database containing all known synthesized structures combining published literature, ICSD, and Materials Project entries result in our most synthesizable structures being already synthesized. Additional unification of this database, as well as separation of whether such structures have been made using solid-state or hydrothermal methods, and under ambient or high pressures would enable the development of highly accurate solid-state synthesizability score. This is especially important as solid-state methods make up the bulk of industrial processing for inorganic compounds. \red{Accordingly, we treat the Materials Project “theoretical” designation as our proxy for a discovery setting.}

\section{Conclusion}

Crystal generation could enable a paradigm shift in the development of novel materials, but judging the plausibility of structures is increasingly important as the number of generated plausible structures increases. Our synthesizability-guided pipeline for materials discovery was able to reliably identify both novel materials and recover materials mislabeled as theoretical in our datasets. Of the 16 targets, we successfully synthesized seven, formed two polymorphs, and failed in the remaining seven cases. Our high success rate (44\%) was achieved with only one experiment per target. 

To our knowledge, this work is among the first demonstrations in which machine-learned synthesizability predictions guided experimental target selection and resulted in previously unreported compounds. Starting from a corpus of 4.4 million candidate structures, our integrated screening prioritized candidates with high predicted synthesizability. Focusing effort on these targets minimized time spent on unrealistic compositions and accelerated laboratory progress toward highest priority materials. Together, these results underscore the practical impact of synthesizability-aware screening in directing material discovery.

\paragraph{Acknowledgments}

The author T.P. wishes to express his sincere gratitude to the Bavarian Ministry of Economic Affairs, Regional Development and Energy and TUMint.Energy Research GmbH for their generous support.

\small
\bibliographystyle{unsrtnat}
\bibliography{refs}

\newpage
\appendix

\section{Technical Appendices and Supplementary Material}

\begin{table}[H]
\centering
\small
\begin{tabular}{ll}
\toprule
\textbf{Symbol} & \textbf{Meaning / Type} \\
\midrule
$x_c$ & Composition (stoichiometric descriptor/vector over elements) \\
$x_s$ & Relaxed crystal structure (atomic positions + lattice); input structure \\
$x=(x_c,x_s)$ & Full input for a candidate material \\
$y\in\{0,1\}$ & Label: $1$ if structure is reported synthesized; $0$ otherwise \\
$f_c(\cdot;\theta_c)$ & Composition encoder (MTEncoder); parameters $\theta_c$ \\
$f_s(\cdot;\theta_s)$ & Structure encoder (GNN fine-tuned from JMP); parameters $\theta_s$ \\
$\mathbf{z}_c$ & Composition embedding $=f_c(x_c;\theta_c)\in\mathbb{R}^{d_c}$ \\
$\mathbf{z}_s$ & Structure embedding $=f_s(x_s;\theta_s)\in\mathbb{R}^{d_s}$ \\
$s_c(x)$ & Synthesizability probability from the composition head, in $[0,1]$ \\
$s_s(x)$ & Synthesizability probability from the structure head, in $[0,1]$ \\

\bottomrule
\end{tabular}
\caption{Notation used in the ML formulation and evaluation.}
\end{table}
\footnotetext{Here $s(i)$ denotes the model probability used in the loss (e.g., per-head); later ranking uses $\mathrm{RankAvg}$.}

\subsection{Classification Metrics}
\label{app:metrics}

We cast synthesizability prediction as a class-imbalanced binary classification task with definitive positive and negative labels. Let $\mathrm{TP},\mathrm{FP},\mathrm{TN},\mathrm{FN}$ denote true/false positives/negatives computed at a decision threshold $\tau$ on the predicted probability $s(x)\in[0,1]$.

\paragraph{Thresholded metrics.}
We report precision, recall (a.k.a.\ true positive rate), and their harmonic mean $F_1$:
\[
\mathrm{Precision}=\frac{\mathrm{TP}}{\mathrm{TP}+\mathrm{FP}},\qquad
\mathrm{Recall}=\frac{\mathrm{TP}}{\mathrm{TP}+\mathrm{FN}},\qquad
F_1=\frac{2\,\mathrm{Precision}\cdot \mathrm{Recall}}{\mathrm{Precision}+\mathrm{Recall}}.
\]

\subsection{Additional Synthesizability Prediction Results}

\begin{figure}[H]
  \centering
  \includegraphics[width=0.7\linewidth]{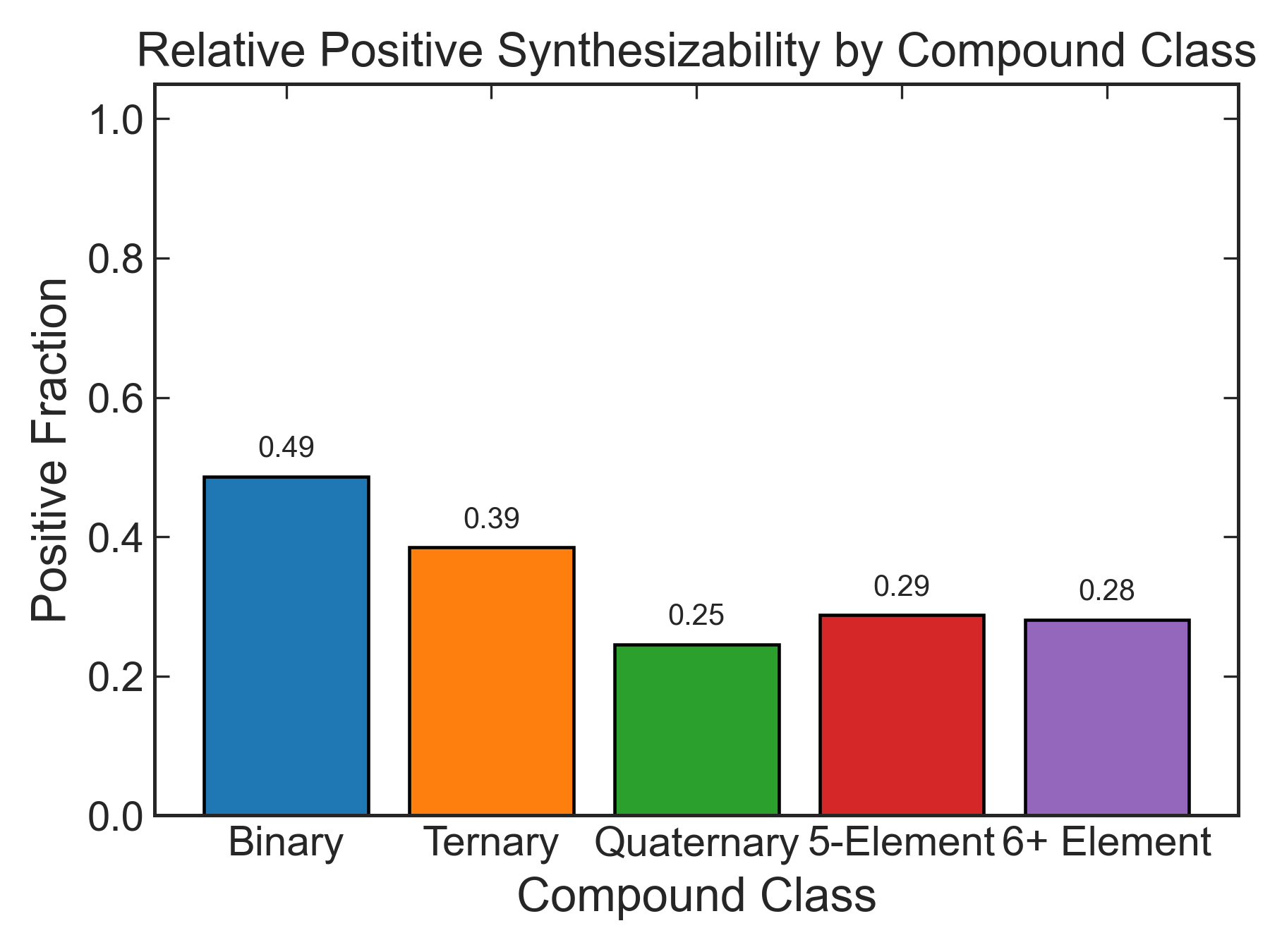}
    \caption{Synthesizability by compound class. Computed from entries in the Materials Project.}
  \label{fig:synthesizability_by_compound_class}
\end{figure}

\subsection{Additional Experimental Details}

The experiments were conducted in our high-throughput laboratory. The proportions of precursor materials were selected to match stoichiometry without consideration for the differing volatility of different precursors. As some precursors are more volatile and therefore disproportionately more 'lost' during the reaction, some purity is likely lost in this non-optimized process. All precursors were combined together and milled for 10 minutes using a Hauschild Speedmixer Smart Dac 400.3 FVZ to reduce particle size and thoroughly mix the components. Subsequently, the powders were placed in alumina crucibles and placed in a Thermo Scientific Thermolyne Benchtop Muffle Furnace for a specified time at a specified temperature. The precursors, times, and temperatures are listed below in Fig. \ref{fig:recipes}. The samples were subsequently characterized using a Malvern Panalytical Aeris benchtop X-ray diffractometer. The fitting of the XRD spectra was done using an in-house code package that uses a BGMN Rietveld refinement as a final step \cite{doebelin2015profex}. The exact fitting procedure is discussed in the Appendix.

Originally, there were 24 targets, of which 8 samples, while fully undergoing synthesis, were not able to be extracted out of their crucibles for XRD analysis. These samples will subsequently be processed and evaluated using the same pipeline, and as such, we do not consider them failed attempts, but rather attempts in progress.

\begin{figure}[!t]
  \centering
  \includegraphics[width=0.7\linewidth]{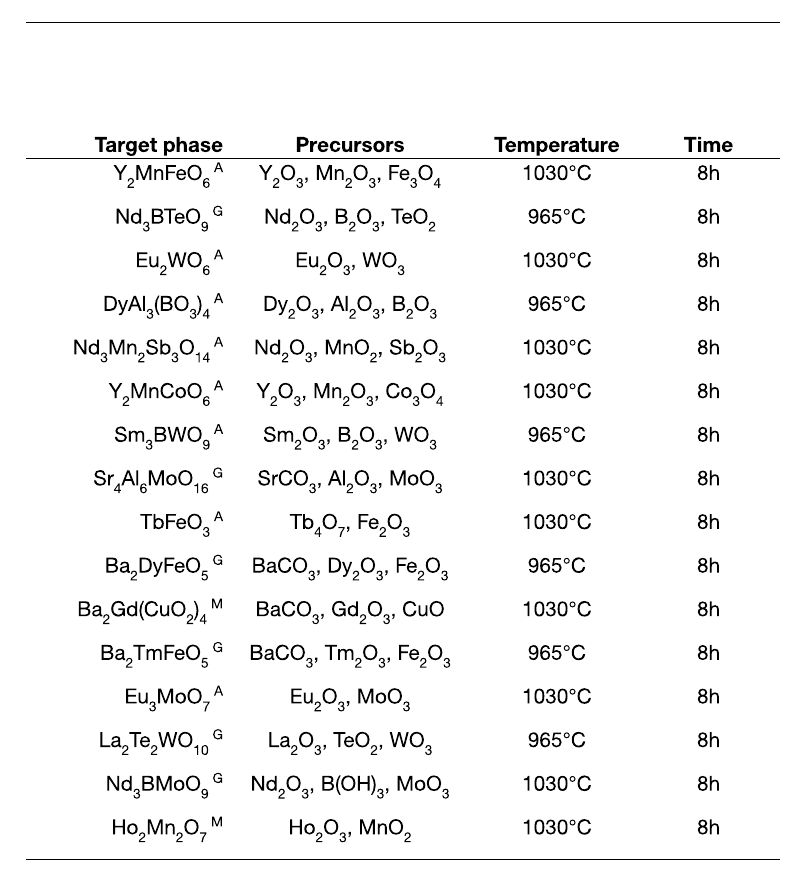}
    \caption{Experimental details for the 16 targets, including precursors, processing temperatures, and times.}
  \label{fig:recipes}
\end{figure}

\newpage
\subsection{XRD Fits for Experiments}

To perform the XRD fits, we used an in-house phase-identification algorithm that picked likely phases from the ICSD, Materials Project, GNoME, and Alexandria. These were weighted based on thermodynamic likelihood. These were then human-validated, before an automated Rietveld refinement procedure was performed using BGMN \citep{doebelin2015profex}, a robust automated Rietveld refinement algorithm. These results were compared against alternative models with different phases to perform an additional qualitative assessment of the model fit. 

We provide XRD fits for all of the claimed successful syntheses below.

\textbf{Y$_2$MnFeO$_6$:}

\begin{figure}[H]
  \centering
  \includegraphics[width=\linewidth]{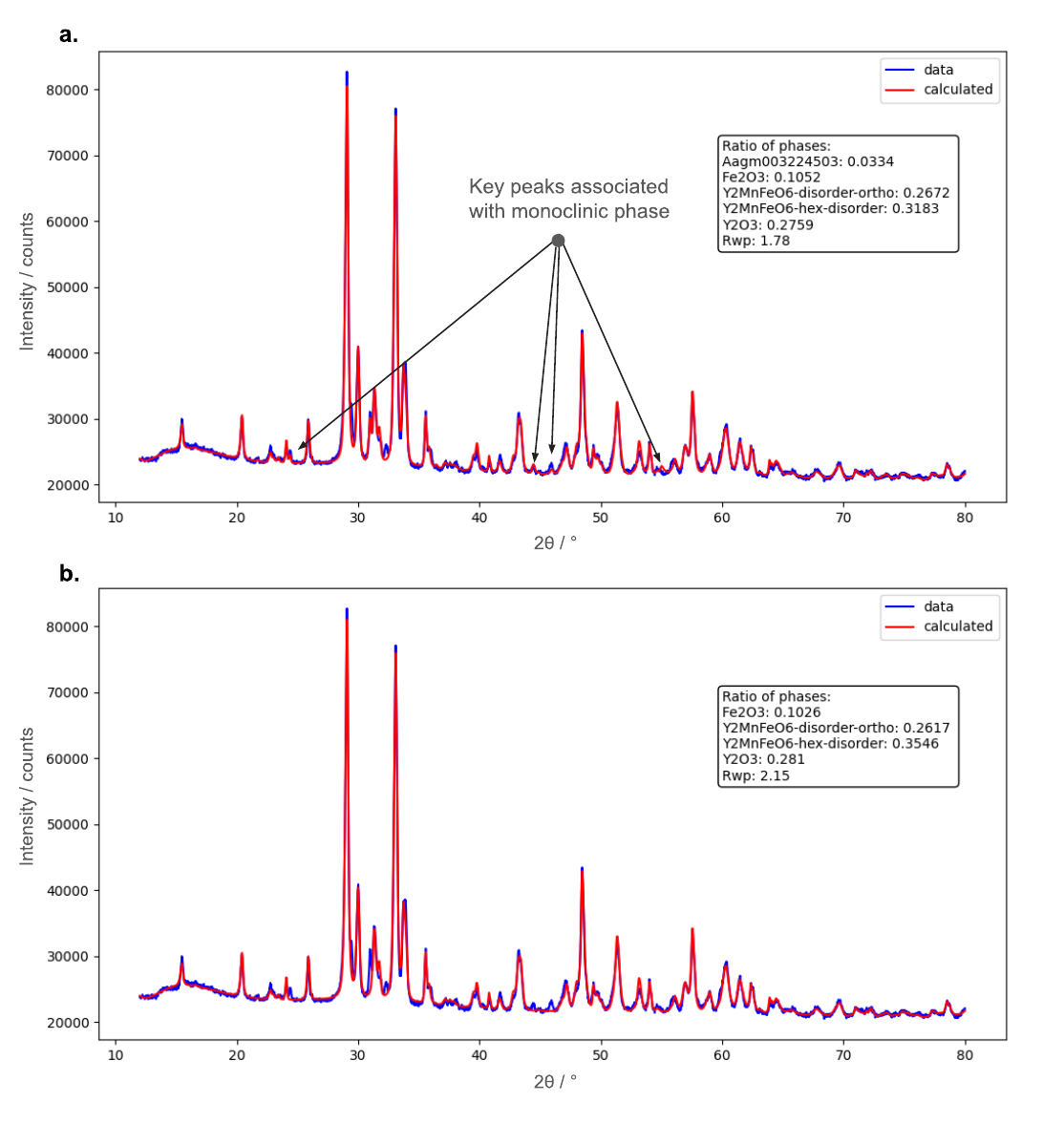}
    \caption{XRD fit for Y$_2$MnFeO$_6$. The small inset rectangle contains phases present (Y$_2$O$_3$, Fe$_2$O$_3$, and several disordered phases of Y$_2$MnFeO$_6$. \textbf{a.} Fit with the target phase (labeled Aagm003224503) present versus \textbf{b.} without the target phase present. A notable improvement in R$_{wp}$ is observed, as well as multiple previously unfitted peaks being fitted, strongly suggesting the presence of the target structure. Known Y-Mn-Fe-O compounds (including two and three metal compounds) did not fit the target peaks.}
  \label{fig:ymfo}
\end{figure}

\newpage
\textbf{Eu$_2$WO$_6$:}

\begin{figure}[H]
  \centering
  \includegraphics[width=\linewidth]{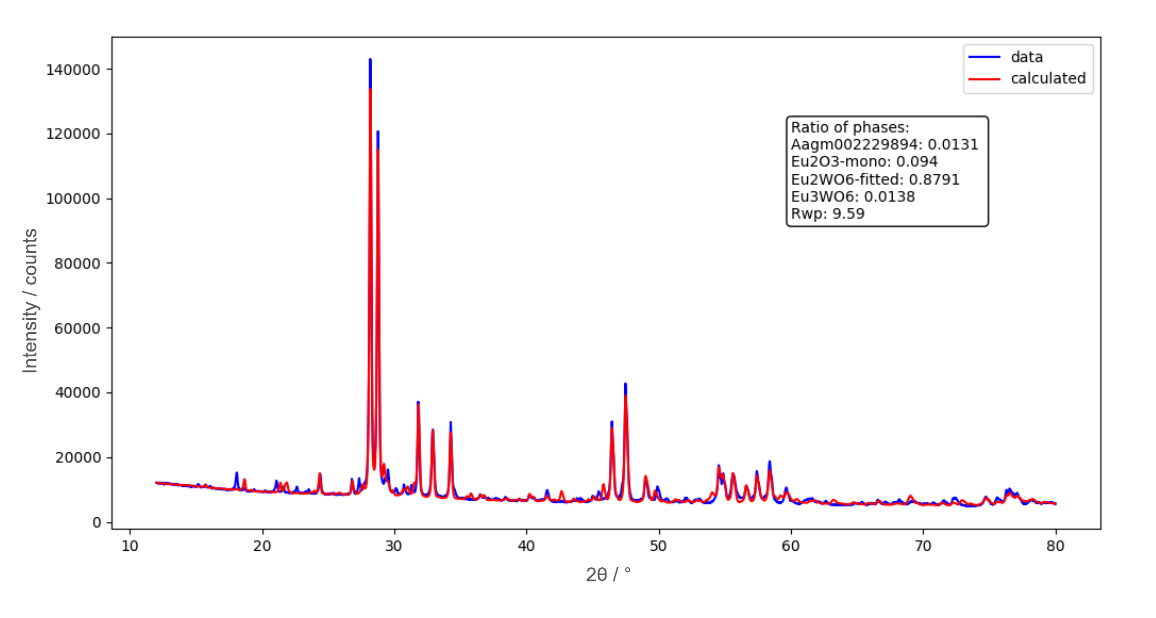}
    \caption{XRD fit for Eu$_2$WO$_6$. The target phase is observed at too low a percentage (<1.5$\%$ to be plausible. Instead a polymorph (88$\%$ purity) with a C2/m with a=14.29 Å, b=3.649 Å, c = 8.876 Å, and $\beta=100.4^\circ$ unit cell is observed instead. Some intensity, likely a complex tungsten oxide, could not be fitted.}
  \label{fig:ewo}
\end{figure}

\textbf{DyAl$_3$(BO$_3$)$_4$:}

\begin{figure}[H]
  \centering
  \includegraphics[width=\linewidth]{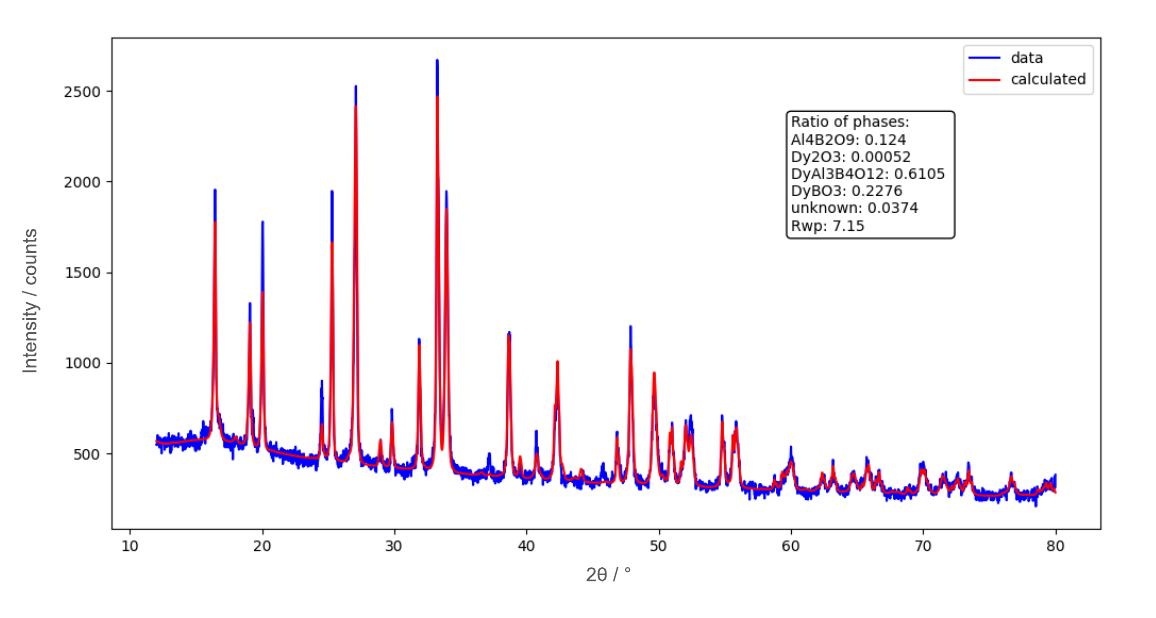}
    \caption{XRD fit for DyAl$_3$(BO$_3$)$_4$. The target phase is observed at 62\%. An unknown orthorhombic phase that does not correspond to known structures was observed at 4\%.}
  \label{fig:dabo}
\end{figure}

\newpage

\textbf{Nd$_3$Mn$_2$Sb$_3$O$_14$:}

\begin{figure}[H]
  \centering
  \includegraphics[width=\linewidth]{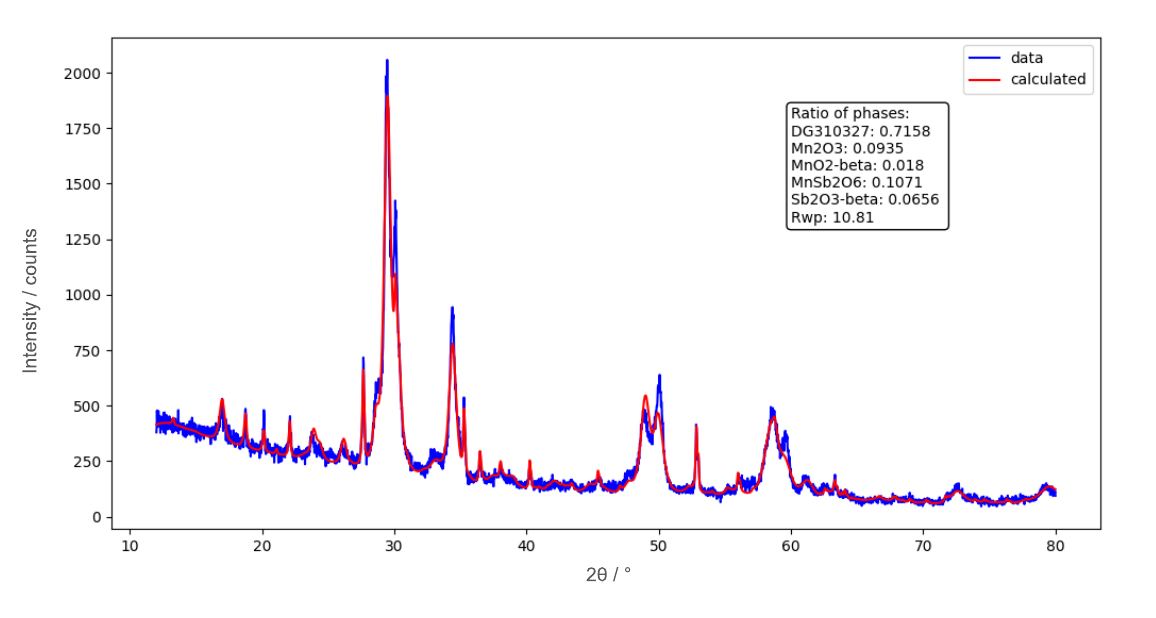}
    \caption{XRD fit for Nd$_3$Mn$_2$Sb$_3$O$_14$. The target phase is listed DG310527 and has 72\% purity.}
  \label{fig:nmso}
\end{figure}

\textbf{Y$_2$MnCoO$_6$:}

\begin{figure}[H]
  \centering
  \includegraphics[width=\linewidth]{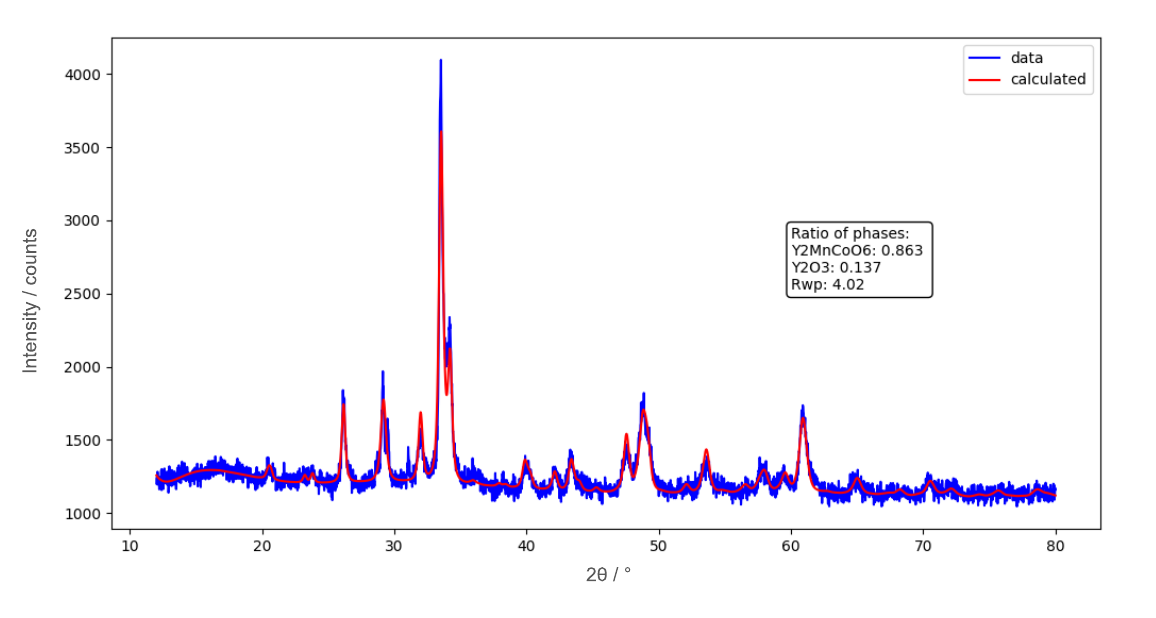}
    \caption{XRD fit for Y$_2$MnCoO$_6$.}
  \label{fig:ymco}
\end{figure}

\newpage

\textbf{Sm$_3$BWO$_9$:}

\begin{figure}[H]
  \centering
  \includegraphics[width=\linewidth]{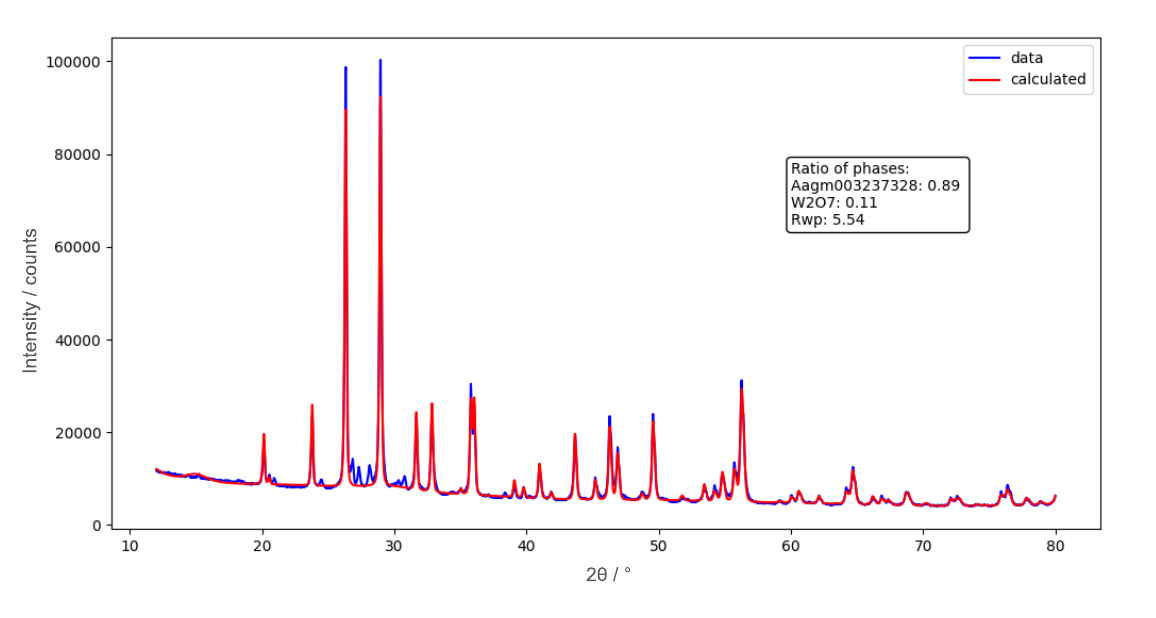}
    \caption{XRD fit for Sm$_3$BWO$_9$. The target phase is listed Aagm003237328.}
  \label{fig:sbwo}
\end{figure}

\textbf{Sr$_4$Al$_6$MoO$_16$:}

\begin{figure}[H]
  \centering
  \includegraphics[width=\linewidth]{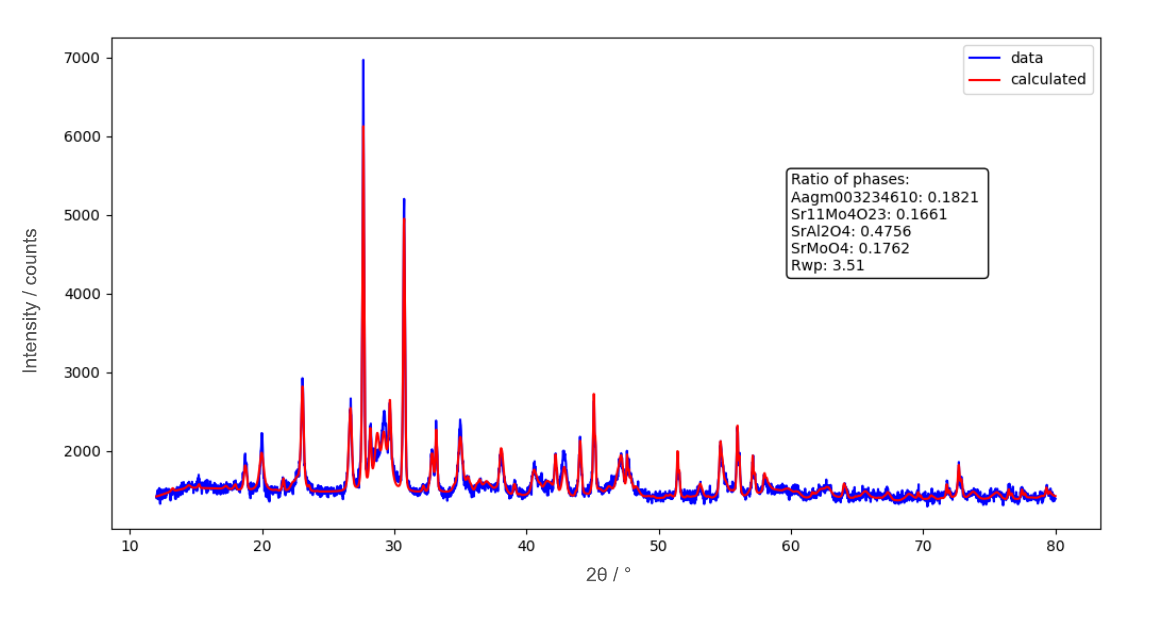}
    \caption{XRD fit for Sr$_4$Al$_6$MoO$_{16}$. The target phase is listed Aagm003234610.}
  \label{fig:samo}
\end{figure}

\newpage

\textbf{TbFeO$_3$:}

\begin{figure}[H]
  \centering
  \includegraphics[width=\linewidth]{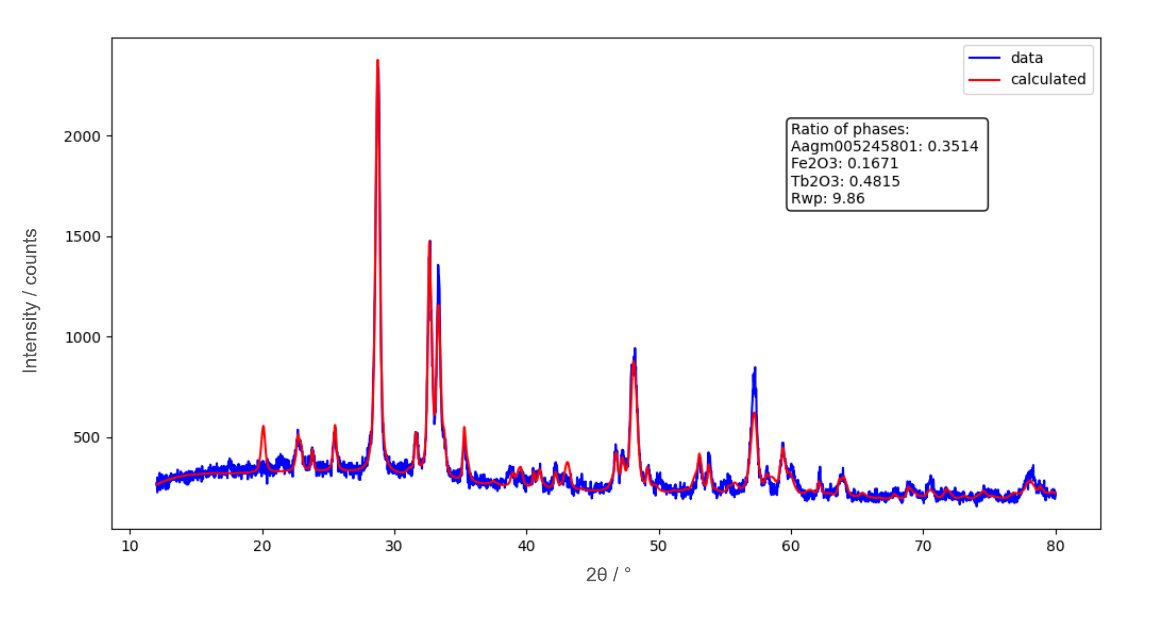}
    \caption{XRD fit for TbFeO$_3$. The target phase is listed Aagm005245801.}
  \label{fig:tfo}
\end{figure}

\textbf{Ba$_2$Gd(CuO$_2$)$_4$:} (provided as an example of a failed synthesis)

\begin{figure}[H]
  \centering
  \includegraphics[width=\linewidth]{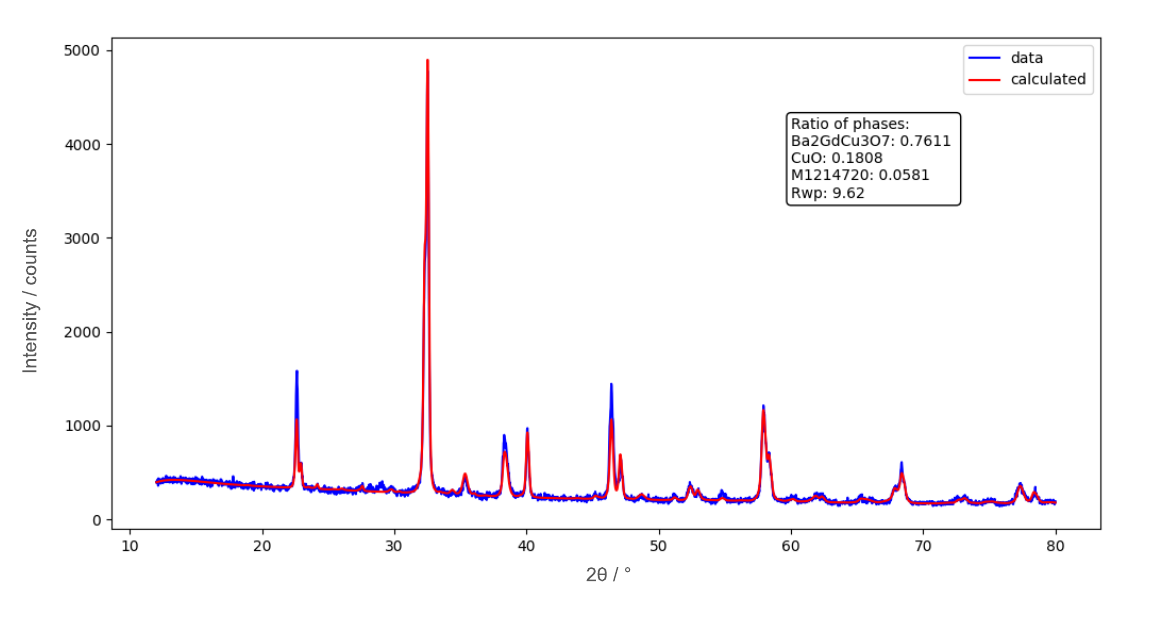}
    \caption{XRD fit for Ba$_2$Gd(CuO$_2$)$_4$. Ba$_2$GdCu$_3$O$_7$ was made instead.}
  \label{fig:bgco}
\end{figure}

\newpage
\subsection{CIF File for Nd$_3$BTeO$_9$}
\begin{verbatim}
data_Nd3BTeO9
_symmetry_space_group_name_H-M   'P 1'
_cell_length_a   8.76075
_cell_length_b   8.76075
_cell_length_c   5.55533
_cell_angle_alpha   90.000
_cell_angle_beta   90.000
_cell_angle_gamma   120.000
_symmetry_Int_Tables_number   1
_chemical_formula_structural   Nd3BTeO9
_chemical_formula_sum   'Nd6 B2 Te2 O18'
_cell_volume   369.25218
_cell_formula_units_Z   2
loop_
 _symmetry_equiv_pos_site_id
 _symmetry_equiv_pos_as_xyz
  1  'x, y, z'
loop_
 _atom_site_type_symbol
 _atom_site_label
 _atom_site_symmetry_multiplicity
 _atom_site_fract_x
 _atom_site_fract_y
 _atom_site_fract_z
 _atom_site_occupancy
  Nd  Nd0  1  0.351800  0.072868  0.204851  1
  Nd  Nd1  1  0.927133  0.278934  0.204851  1
  Nd  Nd2  1  0.721067  0.648200  0.204851  1
  Nd  Nd3  1  0.278934  0.351801  0.704850  1
  Nd  Nd4  1  0.072867  0.721067  0.704850  1
  Nd  Nd5  1  0.648200  0.927133  0.704850  1
  B  B6  1  0.000000  0.000000  0.373877  1
  B  B7  1  0.000000  0.000000  0.873876  1
  Te  Te8  1  0.333334  0.666667  0.246297  1
  Te  Te9  1  0.666666  0.333334  0.746298  1
  O  O10  1  0.537430  0.753814  0.045933  1
  O  O11  1  0.246186  0.783615  0.045933  1
  O  O12  1  0.216386  0.462571  0.045935  1
  O  O13  1  0.129442  0.173277  0.365463  1
  O  O14  1  0.043835  0.870558  0.365463  1
  O  O15  1  0.826722  0.956164  0.365463  1
  O  O16  1  0.132779  0.527094  0.461794  1
  O  O17  1  0.472906  0.605685  0.461794  1
  O  O18  1  0.394315  0.867220  0.461794  1
  O  O19  1  0.753814  0.216385  0.545934  1
  O  O20  1  0.462570  0.246185  0.545934  1
  O  O21  1  0.783614  0.537430  0.545934  1
  O  O22  1  0.173277  0.043835  0.865464  1
  O  O23  1  0.956164  0.129442  0.865464  1
  O  O24  1  0.870558  0.826722  0.865464  1
  O  O25  1  0.605684  0.132779  0.961792  1
  O  O26  1  0.867220  0.472905  0.961792  1
  O  O27  1  0.527094  0.394314  0.961794  1
\end{verbatim}

\end{document}